\documentclass{elsart}

\usepackage{epsfig,pifont}
\begin{document}

\begin{frontmatter}

\title{Directed and elliptic flow in $^{197}$Au + $^{197}$Au at intermediate
energies}

\author[a,j]{J.~{\L}ukasik},
\author[b]{G.~Auger},
\author[a]{M.L.~Begemann-Blaich},
\author[d]{N.~Bellaize},
\author[a]{R.~Bittiger},
\author[d]{F.~Bocage},
\author[c]{B.~Borderie},
\author[d]{R.~Bougault},
\author[b]{B.~Bouriquet},
\author[e]{J.L.~Charvet},
\author[b]{A.~Chbihi},
\author[e]{R.~Dayras},
\author[d]{D.~Durand},
\author[b]{J.D.~Frankland},
\author[c,k]{E.~Galichet},
\author[a]{D.~Gourio},
\author[f]{D.~Guinet},
\author[b]{S.~Hudan},
\author[f]{P.~Lautesse},
\author[c]{F.~Lavaud},
\author[a]{A.~Le~F{\`e}vre},
\author[e]{R.~Legrain\thanksref{dec}},
\author[d]{O.~Lopez},
\author[a]{U.~Lynen},
\author[a]{W.F.J.~M{\"u}ller},
\author[e]{L.~Nalpas},
\author[a]{H.~Orth},
\author[c]{E.~Plagnol},
\author[g]{E.~Rosato},
\author[h]{A.~Saija},
\author[a]{C.~Schwarz},
\author[a]{C.~Sfienti},
\author[d]{B.~Tamain},
\author[a]{W.~Trautmann},
\author[i]{A.~Trzci\'{n}ski},
\author[a]{K.~Turz{\'o}},
\author[d]{E.~Vient},
\author[g]{M.~Vigilante},
\author[e]{C.~Volant},
\author[i]{B.~Zwiegli\'{n}ski}
\collab{The INDRA and ALADIN Collaborations}
\thanks[dec]{deceased}

\address[a]{Gesellschaft f{\"u}r Schwerionenforschung mbH, D-64291 Darmstadt,
Germany}
\address[b]{GANIL, CEA et IN2P3-CNRS, F-14076 Caen, France}
\address[c]{Institut de Physique Nucl{\'e}aire, IN2P3-CNRS et Universit{\'e}, F-91406
Orsay, France}
\address[d]{LPC, IN2P3-CNRS/ENSICAEN et Universit{\'e} 
F-14050 Caen, France}
\address[e]{DAPNIA/SPhN, CEA/Saclay, F-91191 Gif sur Yvette, France}
\address[f]{Institut de Physique Nucl{\'e}aire, IN2P3-CNRS et Universit{\'e}, F-69622
Villeurbanne, France}
\address[g]{Dipartimento di Scienze Fisiche e Sezione INFN, Univ. Federico II,
I-80126 Napoli, Italy}
\address[h]{Dipartimento di Fisica dell' Universit{\`a} and INFN, I-95129
Catania, Italy}
\address[i]{A.~So{\l}{}tan Institute for Nuclear Studies, Pl-00681 Warsaw, Poland}
\address[j]{Institute of Nuclear Physics, Pl-31342 Krak{\'o}w, Poland}
\address[k]{Conservatoire National des Arts et M{\'e}tiers, F75141 Paris, 
France}

\begin{abstract}

Directed and elliptic flow for the $^{197}$Au + $^{197}$Au system at incident
energies between 40 and 150 MeV per nucleon  has been measured using the 
INDRA 4$\pi$ multi-detector. For semi-central collisions, the elliptic flow 
of $Z \leq 2$ particles switches from in-plane to out-of-plane enhancement at 
around 100 MeV per nucleon, in good agreement with the result reported 
by the FOPI Collaboration. The directed flow changes sign at a bombarding 
energy between 50 and 60 MeV per nucleon and remains negative at lower 
energies. The conditions for the appearance and possible origins of 
negative flow are discussed.

\end{abstract}

\end{frontmatter}

\vspace{3mm}

Considerable progress has been made recently in determining the equation  of
state of nuclear matter from heavy-ion reaction data \cite{fuchs01,dani02}. A
prominent role among the available observables is played by the  collective
flow as  it is most directly connected to the dynamical evolution of the
reaction  system, including the momentum dependence of nuclear interactions 
and in-medium effects (for reviews see \cite{reisdorf97,herrmann99}).  Very
significant constraints on the possible range of interaction parameters  have
been derived from transverse and elliptic flow variables \cite{dani02}.

For $^{197}$Au + $^{197}$Au collisions, the amplitudes of transverse and 
elliptic collective motion assume their maxima at bombarding energies  of 300
to 400 MeV per nucleon. At these energies, the sign of the  elliptic flow
indicates a preference for emissions perpendicular  to the reaction plane. The
change-of-sign recently observed at ultra-relativistic energies has received
particular interest as it reflects the increasing pressure
buildup in the non-isotropic collision zone \cite{RHIC1,RHIC2}. Toward the
lower incident energies, the  directed flow observables are presumed to be
related to the competition of mean-field and nucleon-nucleon collision 
dynamics and their evolution with the bombarding energy
\cite{bertsch87,magestro00,andronic01prc}. Elliptic flow
has been identified with the collective motion resulting from the rotation 
of the compound system or the expansion of the hot and compressed participant 
zone, possibly modified by the shadowing effect of the colder spectator matter
\cite{wilson90,tsang92,lacey93,wilson95,shen98,andronic01npa}. Also here, at
the intermediate energies, the transition energies at which the flow parameters
change  sign are particularly useful for the comparison with theory. Their
correct prediction requires the cancellation of the competing  momentum
components which is highly sensitive to specific parameters of the  theoretical
description. A more technical advantage is the weak sensitivity of the 
transition energies to the resolution achieved in reconstructing the reaction
plane.

In this Letter, we present results of the flow analysis applied to the data 
for $^{197}$Au + $^{197}$Au collisions at incident energies from 40 to 150 MeV per
nucleon, collected with the 4$\pi$ INDRA multi-detector \cite{Pouthas} and
with beams provided by the heavy-ion synchrotron SIS at GSI. 
With this energy range, the gap is bridged between existing excitation 
functions of collective motion in the relativistic and Fermi-energy domains.
In agreement with the data of other groups, the transition from 
predominantly in-plane to out-of-plane emissions at about 100 MeV per 
nucleon, as reported by the FOPI collaboration \cite{andronic01npa}, 
is confirmed. Directed flow is found to change its sign at a bombarding 
energy below 60 MeV per nucleon \cite{magestro00}, however with parameters 
that are found to depend strongly on the exact method applied and on the 
experimental acceptance. 

Details of the experiment, including the identification and calibration 
procedures have been presented in Refs. \cite{luka02,lefevre04} and  references
given therein. For impact parameter selection, the total transverse energy
$E_{\perp}^{12}$ of light charged particles ($Z\leq2$)  was used as a sorting
variable. The minimum-bias distributions of this  quantity scale in proportion
to the center-of-mass collision energy,  which supports its usefulness as an
indicator of the collision geometry. A maximum impact parameter $b_{\rm max} =
12$ fm $\pm 10\%$ was deduced from the  measured integrated beam intensity and
the target thickness. It corresponds  to the chosen trigger condition of at
least 5 charged particles detected  and, within errors, remains approximately
constant over the covered range of  bombarding energies. With this information,
and assuming the monotonic relation \cite{cavata90} between the impact
parameter and $E_{\perp}^{12}$, the data were sorted
into six impact-parameter bins, each 2 fm wide and, altogether, covering the
full range 0-12 fm. Particles of interest with $Z\leq2$ were excluded
from $E_{\perp}^{12}$ to reduce autocorrelations, a procedure found useful for peripheral collisions. It was further requested that at least 45\% of the 
total charge of the system are detected, a condition used to reject 
peripheral events in which the projectile residue had escaped detection.

The kinetic-energy tensor, constructed from all identified charged particles,
can be regarded as a global observable characterizing the preferential 
directions of the particle and fragment emissions. The flow angle
$\Theta_{\rm flow}$ is defined as the angle between its largest eigenvector and the
beam axis. Its distributions, analyzed with the weight 1/sin$\Theta_{\rm flow}$ 
\cite{dani83}, are found to be mainly a function of centrality, with 
pronounced peaks at finite angles appearing at smaller impact parameters.
The mean value increases from between 3$^\circ$ and 6$^\circ$ for peripheral 
to about 30$^\circ$ for central collisions. The excitation functions 
are fairly flat except for the most central collisions. For $b \leq 2$ fm,
e.g., the weighted mean flow angle increases from about 
15$^\circ$ to 40$^\circ$ for bombarding energies between 40 and 150~MeV per nucleon. 

\begin{figure} [!htb]	
    \leavevmode
    \centering
    \epsfxsize=0.95\linewidth
   \epsffile{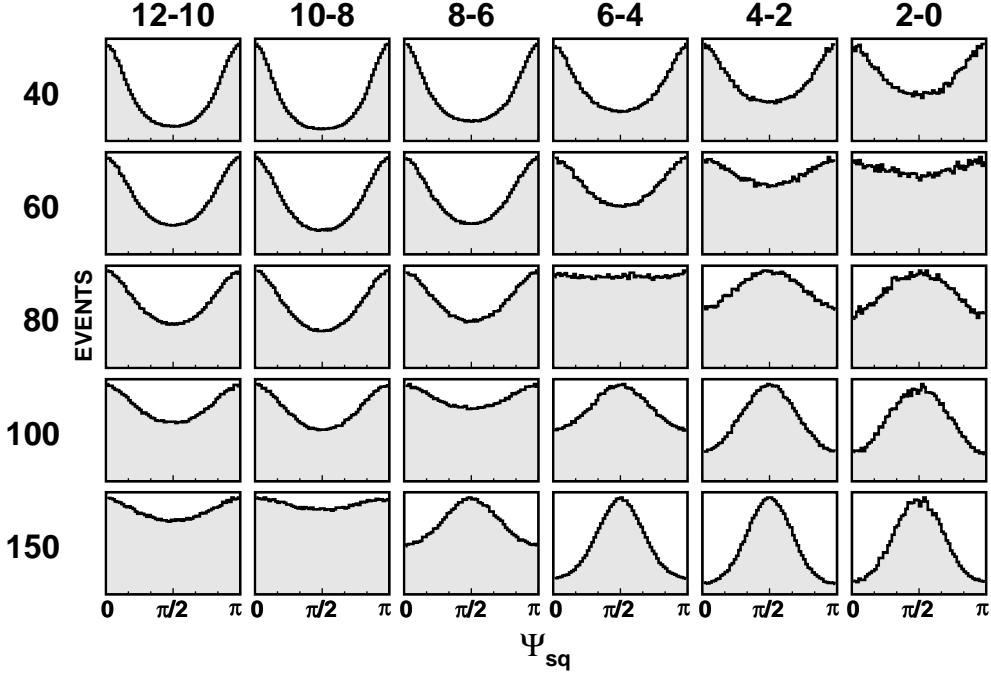}


\caption{{Distributions of the squeeze angle $\Psi_{\rm sq}$ for 
incident energies from 40 to 150~MeV per nucleon (from top to bottom) and 
sorted into 2-fm-wide bins of the deduced impact parameter 
as indicated above each column of panels.}}

\label{fig_sqang} \end{figure}

The squeeze angle $\Psi_{\rm sq}$, defined as the angle between the middle
eigenvector of the kinetic-energy tensor and the reaction plane
\cite{gutbrod90}, characterizes the preferential azimuthal directions of
emission. The squeeze-angle distributions exhibit a
clear trend as a function of incident energy and centrality (Fig.
\ref{fig_sqang}). The  minima at $\pi/2$, observed at lower
energies and more  peripheral impact parameters, indicate predominantly
in-plane emissions. Peaks at $\pi/2$, most strongly pronounced in the more
central bins at the higher incident energies, correspond to a preference for
azimuthal emissions perpendicular to the reaction plane, the so-called
squeeze-out \cite{gutbrod90}.

The curvature of the distributions has been analyzed using their
standard deviation $\sigma(\Psi_{\rm sq})$. The expression 
$\sigma(\Psi_{\rm sq})/(\pi/\sqrt{12})-1$ is positive for concave distributions
of $\Psi_{\rm sq}$, negative for convex distributions, and zero for flat ones. 
The transition energies, $E_{\rm tran}^{\rm sq}$, 
identified by a change-of-sign of this variable, 
have been determined by a linear interpolation between bombarding energies. 
Linear extrapolations were used to obtain estimates for the very
peripheral impact parameter bins.

\begin{figure} [htb]	
    \leavevmode
    \centering
    \epsfxsize=0.95\linewidth
   \epsffile{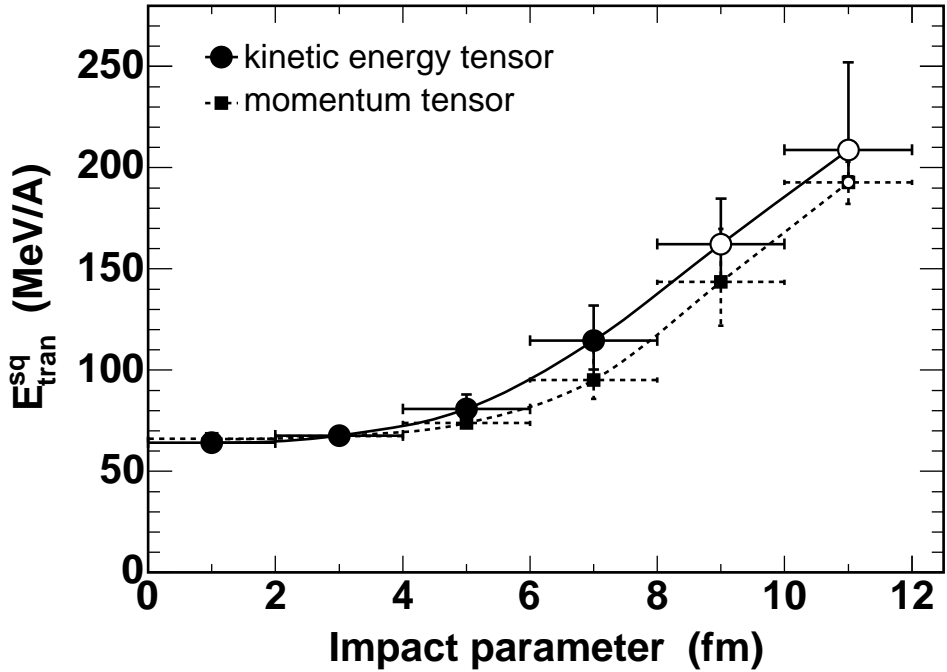}

\caption{{In-plane to out-of-plane transition energies, $E_{\rm tran}^{\rm sq}$,
determined from the curvatures of the squeeze-angle distributions, as a
function of the impact parameter $b$. The full and empty symbols indicate
interpolated and extrapolated values, respectively. The full (dashed) line
connects results obtained from analyses with a kinetic-energy (momentum) tensor
for the event description. The horizontal error bars represent the widths of
the impact-parameter bins and the vertical errors result from the systematic
uncertainty of $b_{\rm max}$.}}

\label{fig_esqang} 
\end{figure}

The energies of the transition from predominantly 
in-plane to out-of-plane
emissions are a strong function of the impact  parameter (Fig.
\ref{fig_esqang}). They extend from 65 MeV per nucleon for central collisions
up to about 200 MeV for the most peripheral collisions and exhibit an
increasingly rapid rise toward the more peripheral impact parameters. Results
obtained by representing the event with the kinetic-energy tensor or with a
momentum tensor, i.e. without the weight factor $1/2m$, are identical at
central impact parameters but diverge slightly in the peripheral bins.

\begin{figure} [!htb]	
    \leavevmode
    \centering
    \epsfxsize=0.95\linewidth
   \epsffile{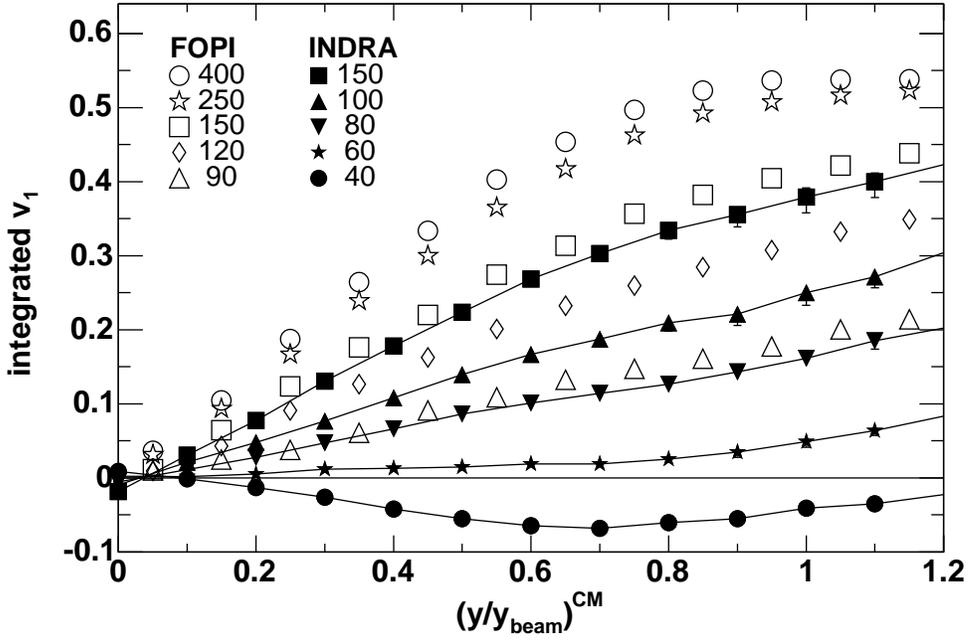}

  \caption{Transverse-flow parameter $v_{1}$ for $Z=2$ particles integrated 
  over $p_{\rm T}$ as a function of
  scaled center-of-mass rapidity for mid-central collisions (2--5.3 fm).  The
  open and filled symbols represent the FOPI \protect\cite{andronic01prc} 
  and the present data, respectively. 
  The numbers in the legend indicate the beam 
  energies per nucleon in MeV. The vertical error bars result from the 
  systematic uncertainty of $b_{\rm max}$.}

\label{fig:v1}
\end{figure}

The diagonalized tensors provide a global representation of the collective
event properties. A much more detailed quantitative information can be obtained
from the Fourier analysis of azimuthal distributions of the reaction products
measured with respect to the reconstructed reaction plane and as a function of
particle type, rapidity $y$ and possibly the transverse momentum $p_{\rm T}$. 
The first two coefficients, $v_{1}$ and $v_{2}$, of the Fourier expansion
characterize the directed and elliptic flow, respectively
\cite{volo96,ollitrault97,poskanzer98,borghini02}. The reaction plane
has been determined by several methods, including the flow-tensor method
\cite{gyulassy82}, the flow Q-vector method \cite{dani85}, and the
azimuthal-correlation method \cite{wilson92}.

The rapidity dependence of the directed-flow 
parameter $v_{1}$ for $Z=2$ particles,
integrated over transverse momentum, is shown in Fig. \ref{fig:v1}. The 
present INDRA data are combined with the FOPI data \cite{andronic01prc},
both measured for mid-central collisions with impact parameters of 2--5.3 fm
and shown without corrections for the  reaction plane dispersion.
In the case of the INDRA
data, the reaction plane has been reconstructed using the Q-vector method
with the weights $\omega=p_{z}$, excluding the particle of interest 
(``1 plane per particle'') and correcting for the effects of momentum 
conservation \cite{ogi89}. 

The slope of $v_1$ as a function of rapidity is seen to rise monotonically 
with energy over the full range of 40 to 400 MeV per nucleon which is 
covered by the two experiments. Good agreement is observed in the overlap 
region which, e.g., may be verified at the incident energy of 150 MeV per 
nucleon which was used in both experiments. The coefficient $v_1$ and its 
slope as a function of rapidity are practically zero at 60 MeV per nucleon 
and become even negative at 40 MeV per nucleon. This intriguing observation 
of a negative flow has already been reported for the lighter systems 
$^{40}$Ar + $^{58}$Ni, $^{58}$Ni + $^{58}$Ni, and $^{129}$Xe + $^{\rm nat}$Sn, 
provided the 1-plane-per-particle method was used \cite{cussol02}. 
For these systems, a balance energy, $E_{\rm bal}$, has been 
determined by associating it with the minima of the approximately parabolic 
excitation functions of the flow parameter which, in the cases of 
$^{40}$Ar + $^{58}$Ni and $^{58}$Ni + $^{58}$Ni, appeared at negative flow values. 
If the same parabolic scenario is adopted for the present case of 
$^{197}$Au + $^{197}$Au, the balance energy should fall below the observed 
zero-crossing of the slope d$v_1$/d$y$ for which the value $54\pm4$~MeV 
per nucleon has been obtained from interpolation.


Values for the balance energy in $^{197}$Au + $^{197}$Au have previously been 
determined by extrapolating from  higher energies 
\cite{zhang90,partlan95,crochet97}, and also by searching for the 
minimum of flow \cite{magestro00}. The extrapolations yielded values 
between 47 and 56~MeV per nucleon with a moderate precision but, 
nevertheless, consistent with the zero-crossing at $54\pm4$~MeV 
per nucleon observed here. The excitation function of flow 
reported in Ref. \cite{magestro00} has a minimum at $E_{\rm bal} = 42 \pm 4$~MeV 
per nucleon but the measured slopes were 
exclusively positive for products with $Z>1$, which is contrary to the 
present data.

\begin{figure} [!htb]	
    \leavevmode
    \centering
    \epsfxsize=0.65\linewidth
   \epsffile{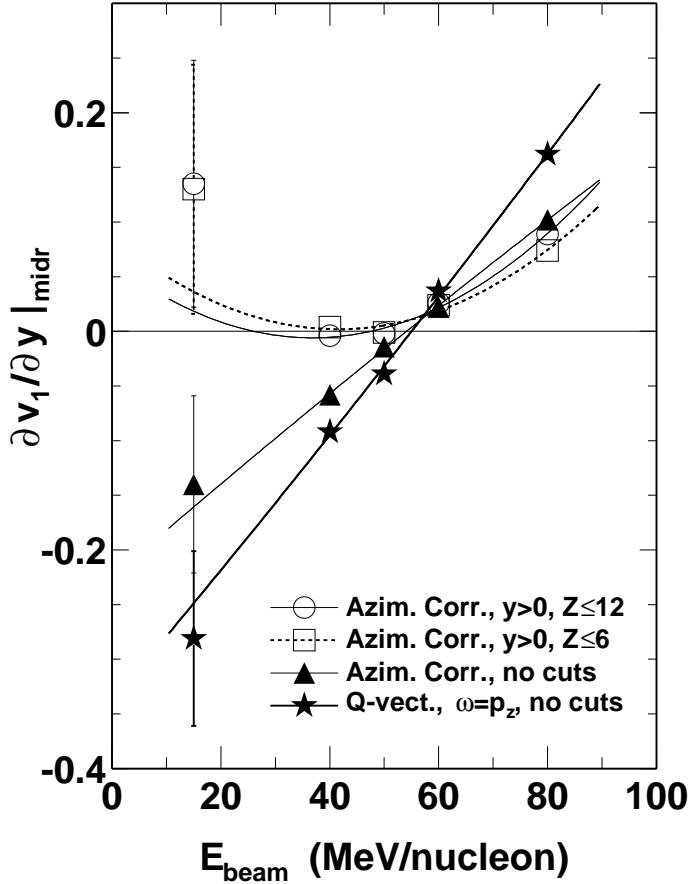}

  \caption{Excitation functions of the mid-rapidity slopes of the $v_{1}$
  parameter for $Z=2$ particles and impact parameters 0-4~fm. 
  The symbols correspond to the indicated methods and conditions
  used to extract the flow values. The lines represent parabolic fits. 
  The statistical errors, only shown for 15~MeV per nucleon, are smaller than 
  the symbol size at other energies.
  }

\vspace{-1mm}

\label{fig:test}
\end{figure}

In order to identify the possible sources of the apparent disagreement 
between these two measurements, the present data has been subjected to 
similar thresholds and selection criteria as used in Ref.~\cite{magestro00} 
and analyzed with the same azimuthal-correlation method of 
reaction-plane reconstruction \cite{wilson92},
excluding the particle of interest. 
The flow is represented by the slopes d$v_1$/d$y$, determined by linear 
fits within the range of the scaled center-of-mass rapidity 
$-0.5 \leq y_{\rm cm}/y_{\rm cm}^{\rm proj} \leq 0.5$. 
The results are summarized in Fig.~\ref{fig:test}, 
including data for 15~MeV per nucleon from a small data sample 
primarily collected for calibration purposes.

With the conditions $Z \leq 12$ or $Z \leq 6$ and by restricting the data to 
positive center-of-mass rapidities, the parabolic excitation function of 
the flow parameter reported in Ref. \cite{magestro00} is qualitatively 
reproduced (open symbols in Fig.~\ref{fig:test}). The latter condition 
has been identified as causing the sizable negative offsets of $v_1$ at 
mid-rapidity in this data. Without these restrictions, the minimum 
disappears, and the flow continues to decrease to more negative 
values, apparently down to bombarding energies as low as 15~MeV per nucleon. 
The Q-vector method 
of the reaction plane
reconstruction with the weights $\omega=p_{z}$ yields larger absolute
flow values. The larger weights assigned to the faster and heavier fragments 
may cause a better definition of the reaction plane 
\cite{ogilvie89_1}.

\begin{figure} [!htb]	
    \leavevmode
    \centering
     \epsfxsize=\columnwidth
   \epsffile{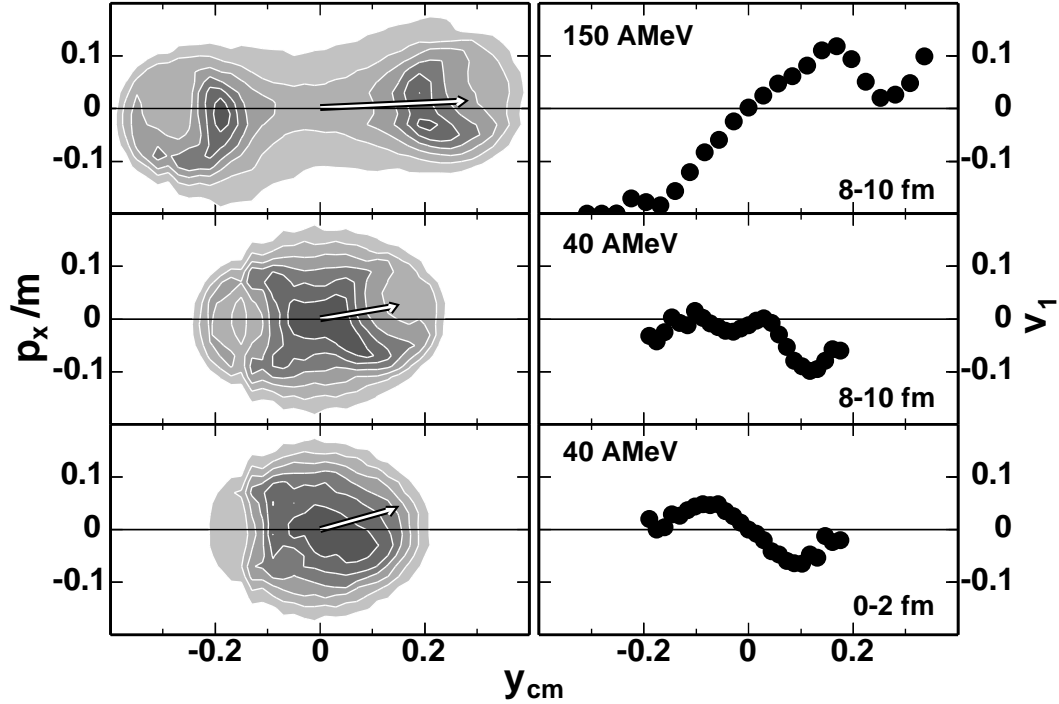}

  \caption{Contour plots (in linear scale) 
  of the in-plane component of the transverse 
  velocity ($p_{x}/m$) versus the center-of-mass rapidity $y_{\rm cm}$ (left column)
  and the mean value $v_1$ as a function of $y_{\rm cm}$ (right column) for $Z=2$ 
  particles and three selected cases of incident energy and impact parameter 
  as indicated. 
  The arrows represent the directions of the three-dimensional
  Q-vector. 
  } 

  \vspace{-1mm}

\label{fig:proj}
\end{figure}

The origin of the negative directed flow observed in the present data and the 
strong effects of the acceptance and selection criteria are illustrated in 
Fig.~\ref{fig:proj}. The left row of panels shows contour plots of the 
in-plane transverse velocity versus the center-of-mass rapidity. 
For peripheral collisions at high incident energy (top panels) the 
deflections of the projectile and target, as represented by the 
three-dimensional Q-vector, are small. The Coulomb repulsion from the 
heavy residues leads to the apparent depression of helium yields near 
the entrance-channel rapidities of $\pm$0.28 and to maximum intensities at 
lower absolute rapidity, as discussed in Ref.~\cite{luka03}. The stronger 
deflection of mid-rapidity particles is evident from the rapid rise of 
$v_1$ with $y_{\rm cm}$ which, at $y_{\rm cm} \approx 0.2$, starts to be modified 
by the effect of the projectile and target spectators.

At 40~MeV per nucleon, and the same peripheral impact-parameters  (middle
panels of Fig.~\ref{fig:proj}), the structure of $v_1$ as a function of $y_{\rm
cm}$ is qualitatively the same as at 150~MeV per nucleon but  compressed into a
smaller range of absolute rapidity. In central collisions,  at this energy, the
distributions are even more compact  (Fig. 5, bottom left). The flow vector
indicates the mean transverse  deflection of the forward-emitted part of the
event. This concentration of  mass and charge, apparently, causes the $Z = 2$
particles to be  preferentially deflected to the opposite side. The resulting
flow,  evaluated around mid-rapidity, is negative.  Clearly, Q-vectors that do
not include the heavier fragments causing this deflection or shadowing  will
have different directions, and the flow measured relative to them may  appear
positive.

The absolute sign of the deflection angle cannot be determined with the 
present methods. There is little doubt, however, that the positive flow 
above 60 MeV per nucleon is associated with an overall positive deflection 
as a result of the dominant nucleon-nucleon dynamics.
Near and above the Coulomb barrier, positive deflection angles follow 
from the dominance of the Coulomb forces.
The present 
observation of a negative flow below 60 MeV per nucleon may thus indeed 
indicate a transition toward predominantly negative-angle emissions
for light products, as concluded in Ref.~\cite{magestro00}, but also 
consistent with polarization measurements for light fragments, 
even though for other reactions \cite{tsang88}.
Thus, the commonly invoked picture of the attractive mean field globally
balancing the repulsive effect of the collisions seems be too simple 
here to explain the observed sign change of flow. The main difference, 
as compared to lighter reaction systems, 
is the enlarged Coulomb field which not only has
a strong impact on the entrance and exit channel trajectories \cite{soff95} 
but also manifests itself in large repulsion effects (larger so-called 
Coulomb rings).

\begin{figure} [!htb]	
    \leavevmode
    \centering
     \epsfxsize=\columnwidth
   \epsffile{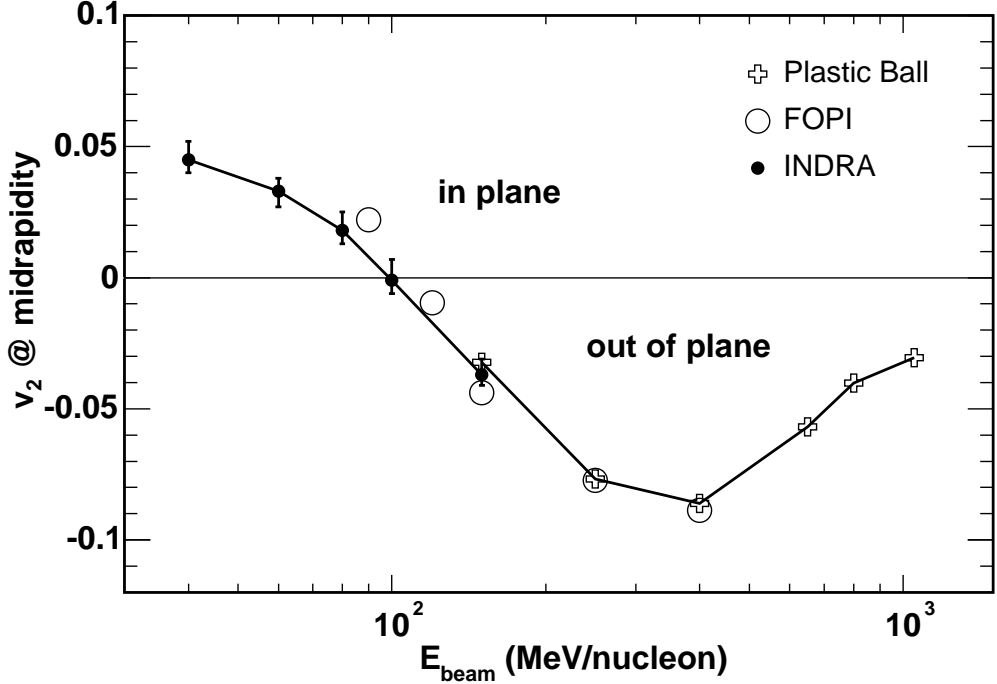}

  \caption{Elliptic-flow parameter $v_{2}$ at mid-rapidity for $Z\leq2$ 
  particles from mid-central collisions, in the rotated reference frame.
  The dots, circles, and crosses represent the INDRA, the FOPI 
  \protect\cite{andronic01npa}, and the Plastic Ball \protect\cite{gutbrod90} 
  data, respectively. The errors shown for the INDRA data are mainly 
  systematic and caused by the uncertainty of $b_{\rm max}$.}

\vspace{-1mm}

\label{fig:fig6}
\end{figure}


The elliptic-flow parameter $v_2$ permits a detailed study of azimuthal 
emissions including their mass and $p_{\rm T}$ dependence, 
as reported for incident energies of 
90~MeV per nucleon and higher in Ref.~\cite{andronic01npa}.
The present data set, due to a higher efficiency for fragments, also reveals
a significant effect of the fragment charge. Over the range of atomic 
numbers $Z \leq 2$ to $Z = 9$, the transition energy $E_{\rm tran}$,
evaluated in the laboratory frame,
decreases from about 100 to 65~MeV per nucleon. The restriction
to light particles, on the other hand, permits a comparison with other data
sets and an extension of the measured excitation function deep into the 
relativistic regime. 

The excitation function of the elliptic flow for $Z\leq2$ particles,
in the rotated reference frame and for semi-central collisions of 4--6 fm,
as measured by the Plastic Ball \cite{gutbrod90}), 
the FOPI \cite{andronic01npa}, and the INDRA collaborations,
is shown in Fig.~\ref{fig:fig6}. For this purpose, the present data
has been analyzed with the same method as the Plastic Ball data, i.e. 
by using the kinetic-energy tensor for the reconstruction of the reaction 
plane and excluding the particle of interest. None of the
three data sets were corrected for the reaction-plane dispersion. This
correction is not expected to substantially alter the value of $E_{\rm tran}$ 
(cf. Ref.~\cite{andronic01npa}) but will affect the values of $v_{2}$ at 
the lower energies at which the reaction plane is less well defined 
because of smaller particle and fragment multiplicities.

The observed agreement between the different experiments is very satisfactory.
In particular, the value of the transition energy of around 100~MeV per
nucleon, obtained by the FOPI 
Collaboration, is confirmed by the present data. 
Because of the $Z$ dependence, it is approximately 20~MeV
higher than the global transition energy derived from the kinetic-energy tensor
for the same impact-parameter bin (Fig.~\ref{fig_esqang}).
Together, the three data sets constitute a coherent systematics of $v_{2}$
which can be expected to serve as a valuable constraint for transport models.


In summary, the presented analysis of $^{197}$Au + $^{197}$Au 
collisions at intermediate energies 
extends the existing excitation functions of the flow parameters 
$v_1$ and $v_2$ from the relativistic regime down to the Fermi-energy regime. 
The observed transition from predominantly in-plane to out-of-plane 
emissions at about 100 MeV per nucleon for $Z \leq 2$ particles
has been confirmed. The directed transverse flow 
changes its sign below 60~MeV per nucleon and becomes
increasingly negative at lower bombarding energies. 
Negative flow and the absence of a parabolic excitation function of
directed flow in $^{197}$Au + $^{197}$Au 
were shown to be connected to the large Coulomb effects in this
reaction system. 

The authors would like to thank A. Andronic and W. Reisdorf for making their
data available and for valuable discussions. M.B. and C.Sc. acknowledge the
financial support of the Deutsche  Forschungsgemeinschaft under the Contract
No. Be1634/1-1 and Schw510/2-1,  respectively; D.Go. and C.Sf. acknowledge the
receipt of  Alexander-von-Humboldt fellowships. This work was supported by the
European Community under contract ERBFMGECT950083.


\begin{thebibliography}{99}
\itemsep -1pt 

\bibitem{fuchs01}
C.~Fuchs et al.,
Phys. Rev. Lett. 86 (2001) 1974.

\bibitem{dani02} 
P.~Danielewicz et al., 
Science 298 (2002) 1592.

\bibitem{reisdorf97} 
W.~Reisdorf and H.G.~Ritter, 
Annu. Rev. Nucl. Part. Sci. 47 (1997) 663.

\bibitem{herrmann99}
N.~Herrmann et al.,  
Annu. Rev. Nucl. Part. Sci. 49 (1999) 581.

\bibitem{RHIC1}
C.~Adler et al., 
Phys. Rev. Lett. 90 (2003) 032301.

\bibitem{RHIC2}
C.~Adler et al., 
Phys. Rev. Lett. 91 (2003) 182301.

\bibitem{bertsch87}
G.F.~Bertsch, W.G. Lynch, M.B. Tsang,
Phys. Lett. B 189 (1987) 384.

\bibitem{magestro00} 
D.J.~Magestro et al., 
Phys. Rev. C 61 (2000) 021602.

\bibitem{andronic01prc} 
A.~Andronic et al., 
Phys. Rev. C 64 (2001) 041604.

\bibitem{wilson90}
W.K.~Wilson et al., 
Phys. Rev. C 41 (1990) 1881.

\bibitem{tsang92}
M.B.~Tsang et al., 
Phys. Rev. C 47 (1992) 2717.

\bibitem{lacey93}
R.A.~Lacey et al., 
Phys. Rev. Lett. 70 (1993) 1224.

\bibitem{wilson95}
W.K.~Wilson et al., 
Phys. Rev. C 51 (1995) 3136.

\bibitem{shen98}
W.Q.~Shen et al., 
Phys. Rev. C 57 (1998) 1508.

\bibitem{andronic01npa} 
A.~Andronic et al., 
Nucl. Phys. A 679 (2001) 765.

\bibitem{Pouthas} 
J.~Pouthas et al., 
Nucl. Instrum. Methods Phys. Res. A 357 (1995) 418.

\bibitem{luka02}
J.~{\L}{}ukasik et al.,
Phys. Rev. C 66 (2002) 064606.

\bibitem{lefevre04}
A.~Le~F{\`e}vre et al.,
Nucl. Phys. A 735 (2004) 219.

\bibitem{cavata90}
C.~Cavata et al.,
Phys. Rev. C 42 (1990) 1760.

\bibitem{dani83}
P.~Danielewicz and M. Gyulassy,
Phys. Lett. 129B (1983) 283.

\bibitem{gutbrod90} 
H.H.~Gutbrod et al., 
Phys. Rev. C 42 (1990) 640.

\bibitem{volo96}
S.~Voloshin and Y.~Zhang,
Z. Phys. C 70 (1996) 665.

\bibitem{ollitrault97} 
J.-Y.~Ollitrault, 
preprint nucl\_ex/9711003.

\bibitem{poskanzer98} 
A.M.~Poskanzer and S.A. Voloshin, 
Phys. Rev. C 58 (1998) 1671.

\bibitem{borghini02} 
N.~Borghini et al., 
Phys. Rev. C 66 (2002) 014901.

\bibitem{gyulassy82} 
M. Gyulassy et al., 
Phys. Lett. B110 (1982) 185.

\bibitem{dani85} 
P.~Danielewicz and G.~Odyniec, 
Phys. Lett. B157 (1985) 146.

\bibitem{wilson92}
W.K. Wilson et al., 
Phys. Rev. C 45 (1992) 738.

\bibitem{ogi89} 
C.A.~Ogilvie et al., 
Phys. Rev. C 40 (1989) 2592.

\bibitem{cussol02} 
D.~Cussol et al., 
Phys. Rev. C 65 (2002) 044604.

\bibitem{zhang90} 
W.M.~Zhang et al., 
Phys. Rev. C 42 (1990) 491.

\bibitem{partlan95} 
M.D.~Partlan et al., 
Phys. Rev. Lett. 75 (1995) 2100.

\bibitem{crochet97} 
P.~Crochet et al., 
Nucl. Phys. A 624 (1997) 755.

\bibitem{ogilvie89_1} 
C.A.~Ogilvie et al., 
Phys. Rev. C 40 (1989) 654.

\bibitem{luka03}
J.~{\L}{}ukasik et al.,
Phys. Lett. B 566 (2003) 76.

\bibitem{tsang88} 
M.B.~Tsang et al., 
Phys. Rev. Lett. 60 (1988) 1479.

\bibitem{soff95} 
S.~Soff et al., 
Phys. Rev. C 51 (1995) 3320.


\end{thebibliography}
\end{document}